\documentclass[12pt]{article}
\usepackage{a4,amssymb,verbatim,amsmath}
\numberwithin{equation}{section}

\begin{document}

\def\dx#1{{\partial \over \partial#1}}
\def\dh#1{\mathop {#1}\limits_{h}}
\def\dhp#1{\mathop {#1}\limits_{+h}}
\def\dhm#1{ \mathop{#1}\limits_{-h}}
\def\dphh#1{ \mathop{#1}\limits_{h \bar h}}
\def\da#1{ \mathop{#1}\limits_{+\tau}}
\def\db#1{ \mathop{#1}\limits_{-\tau}}
\def\dc#1{ \mathop{#1}\limits_{\pm \tau}}
\def\dd#1{ \mathop{#1}\limits_{+h}}
\def\df#1{ \mathop{#1}\limits_{-h}}
\def\dpm#1{ \mathop{#1}\limits_{\pm h}}
\def\dg#1{ \mathop{#1}\limits_{\pm h}}
\def\dh#1{ \mathop{#1}\limits_ h}

\def\sso#1{\ensuremath{\mathfrak{#1}}}

\newcommand{\ddt}{\partial \over \partial t}
\newcommand{\ddx}{\partial \over \partial x}
\newcommand{\ddy}{\partial \over \partial y}
\newcommand{\ddyy}{\partial \over \partial y'}

\begin{center}
{\Large {\bf Invariance and first integrals of \\
canonical Hamiltonian equations}}
\end{center}

\begin{center}

{\large  Vladimir Dorodnitsyn}$^{*}$ {\large and Roman
Kozlov}$^{\ddag}$

\medskip
\hspace{1.5 cm}

${}^{*}$ Keldysh Institute of Applied Mathematics,
Russian Academy of Science,\\
Miusskaya Pl.~4, Moscow, 125047, Russia; \\
E-mail address: dorod@spp.Keldysh.ru \\

${}^{\ddag}$ Department of Mathematics, University of Bergen, \\
Johannes Brunsgate~12, 5008 Bergen, Norway; \\
E-mail address: Roman.Kozlov@math.uib.no

\bigskip
\bigskip

  { \bf 15.05.2009}
  \end{center}

\bigskip

\bigskip
\begin{center}
{\bf Abstract}
\end{center}
\begin {quotation}
In this paper we consider the relation between symmetries and first
integrals of canonical Hamiltonian equations. Based on a newly
established identity (which is an analog of well--known Noether's
identity for Lagrangian framework), this approach provides a simple
and clear way to construct first integrals with the help of
symmetries of a Hamiltonian. The approach is illustrated by a number
of examples, including equations of the three-dimensional Kepler
motion.
\end{quotation}

\section{Introduction}

It has been known since E.~Noether's fundamental work that
conservation laws of differential equations are connected with
their symmetry properties~\cite{Noe}. For convenience we present
here some well--known results (see also, for example,
\cite{Abraham, Goldstein, Arnold})
for both Lagrangian and Hamiltonian
approaches to conservation laws (first integrals).

Let us consider the functional
\begin{equation}   \label{lagyr}
\mathbb{L}(u) =  \int_{\Omega} L ( x,u,u_1) dx ,
\end{equation}
where
$ x=(x^1,x^2,...,x^m)$ are independent variables,
$ u=(u^1,u^2,...,u^n)$ are dependent variables,
$  u_1=(u^k_i)$ are all first order derivatives 
$ u^k_i= {\partial u^k \over \partial x^i  }$, 
$\Omega$ is a domain in $ \mathbb{R}^m$
and $L ( x,u,u_1)$ is  {\it a first order} Lagrangian.
The functional (\ref{lagyr}) achieves its extremal values when $u(x)$
satisfies the  Euler--Lagrange equations
\begin{equation}   \label{extyr}
{ \delta L \over \delta u^k } = {\partial L \over \partial u^k  }
- D_i \left( {\partial L \over \partial u^k_i } \right) = 0,
\qquad
k=1,...,n,
\end{equation}
where
$$
D_i = {\ddx^i} + u^k_i { \partial \over\partial u^k } +
u^k_{ji} { \partial \over\partial u^k_j } + \cdots ,
\qquad i = 1,...,m
$$
are total differentiation operators with respect 
to independent variables $x^ i$. Here and below we assume
summation over repeated indexes. Note that equations (\ref{extyr})
are second order PDEs.

We consider a Lie point transformation group $G$ generated by
the infinitesimal operator
\begin{equation}  \label{symmetry0}
X = \xi^i ( x,u)  { \partial \over \partial x^i } + \eta ^k  (x,u)
{ \partial \over \partial u^k } +  ...,
\end{equation}
where dots mean an appropriate prolongation of the operator on partial
derivatives \cite{Ovs, Olver, Ibr, Blu}.
The group $G$ is called a variational symmetry of the functional
$\mathbb{L}(u) $ if and  only if the Lagrangian satisfies~\cite{Noe}
\begin{equation} \label{cond}
  X(L) + L D_i( \xi^i) = 0,
\end{equation}
where  $ X$ is the first prolongation, i.e. the prolongation  of the
vector field $X$ on the first derivatives $u^k_i$. We will actually
need a weaker invariance condition than given by Eq.~(\ref{cond}).
The vector field $X$ is a divergence symmetry of the functional
$\mathbb{L}(u) $ if there exists vector functions $V^i(x,u,u_1)$, 
$i = 1, ...,m$ 
such that \cite{Bess} (see also [6,7])
\begin{equation} \label{cong}
 X(L) + L D_i( \xi^i) = D_i(V^i) .
\end{equation}

An important result for us is the following:
If $X$ is a variational symmetry
 of the functional $\mathbb{L}(u) $, it is also a
symmetry of the corresponding Euler--Lagrange equation. The
symmetry group of Eqs.~(\ref{extyr}) can of course be larger than
the group generated  by  variational and divergence symmetries
of the Lagrangian.

 Noether's theorem \cite{Noe}  states that for a Lagrangian satisfying
the condition (\ref{cond}) there exists a conservation law of the
Euler--Lagrange equations (\ref{extyr}):
\begin{equation} \label{conserv}
D_i \left(\xi^i L + (\eta^k - \xi^j u^k_j){ \partial L\over \partial
u^k_i } \right) = 0.
\end{equation}
This result can be generalized:
If $X$ is a divergence symmetry of the functional $\mathbb{L}(u) $,
i.e. equation  (\ref{cong}) is satisfied, then there exists a
conservation law
\begin{equation} \label{cdrtre}
D_i \left(\xi^i L + (\eta^k - \xi^j u^k_j){ \partial L\over \partial
u^k_i }-V^i \right) = 0
\end{equation}
of the corresponding Euler--Lagrange equations.

The strong version of the Noether's theorem \cite{Ibr} states that
there exists a conservation law of the Euler--Lagrange
Eqs.~(\ref{extyr}) in the form (\ref{conserv}) if and only if
the condition (\ref{cond}) is satisfied on the solutions of
Eqs.~(\ref{extyr}).

\bigskip

In the present paper we  are interested in canonical Hamiltonian
equations
\begin{equation}  \label{canonical}
\dot{q} ^i = { \partial H \over  \partial {p}_i } , \qquad
\dot{p}_i = - { \partial H \over  \partial {q}^i }, \qquad
i = 1, ... , n .
\end{equation}
These equations can be obtained
by the variational principle
from the action functional
\begin{equation}  \label{principe}
\delta   \int_{t_1} ^{t_2}
\left(
 p_i   \dot{q} ^i  - H ( t,  {\bf q} , {\bf p} )
\right) dt = 0
\end{equation}
in the phase space $( {\bf q}, {\bf p} )$,
where ${\bf q} = ( q ^1, q ^2,..., q ^n ) $,
${\bf p} = ( p_1, p_2, ..., p_n ) $ (see, for
example, \cite{Gelfand, Marsden}).
Variations $ \delta q ^i  $ and  $  \delta p_i $ are arbitrary  and satisfy $
\delta q ^i  ( t_1) = \delta  q ^i ( t_2) =  0 $, $ i = 1, ..., n
$. Then,  we have
$$
\delta   \int_{t_1} ^{t_2}
\left( p_i   \dot{q} ^i  - H ( t, {\bf q} , {\bf p} ) \right) dt
= \int_{t_1} ^{t_2} \left(
\delta  p_i  \dot{q} ^i  +  p_i    \delta  \dot{q} ^i
-  { \partial H \over \partial q ^i }  \delta q ^i
-  { \partial H \over \partial p_i }  \delta p_i
\right) dt
$$
$$
=  \int_{t_1} ^{t_2}   \left[
\left(  \dot{q} ^i
-  { \partial H \over \partial p_i } \right) \delta p_i
- \left(  \dot{p}_i +  { \partial H \over \partial q ^i } \right) \delta q ^i
\right] dt
+ \left[    p_i \delta  q ^i \right]_{t_1} ^{t_2}  .
$$
The last term vanishes because $ \delta q ^i = 0 $ at the
endpoints. Since variations $ \delta q ^i $ and $ \delta p_i $ are
arbitrary, the variational principle (\ref{principe}) is
equivalent to the canonical Hamiltonian equations (\ref{canonical}).

Let us note that the canonical Hamiltonian equations
(\ref{canonical}) can be obtained by action of the variational
operators
\begin{equation}  \label{varoperator1}
{ \delta \over \delta p _i }  =
{ \partial \over \partial p_i }
- D { \partial \over \partial \dot{p}_i } ,
\qquad
i = 1, ..., n
\end{equation}
and
\begin{equation}  \label{varoperator2}
{ \delta \over \delta q ^i }  =
{ \partial \over \partial q ^i }
- D { \partial \over \partial \dot{q} ^i } ,
\qquad
i = 1, ..., n  ,
\end{equation}
where
\begin{equation}  \label{derivative}
D = { \partial  \over  \partial t }
+     \dot{q} ^i { \partial \over \partial q ^i }
+    \dot{p}_i { \partial \over  \partial p_i } + ...
\end{equation}
is the operator of total differentiation with respect to time,
on the function
$$
  p_i   \dot{q} ^i  - H ( t, {\bf q} , {\bf p} ) .
$$
The Legendre transformation relates Hamiltonian and Lagrange
functions
\begin{equation}  \label{Legendre}
 L  ( t, {\bf q} , { \dot{\bf q}}  ) = p_i   \dot{q} ^i  -
 H ( t, {\bf q} , {\bf p} ),
\end{equation}
where ${\bf p}={ \partial L\over \partial \dot{\bf q} }, {\dot{\bf
q}}={\partial H\over \partial {\bf p} }$. It makes it possible to
establish the equivalence of the Euler--Lagrange and Hamiltonian
equations \cite{Arnold}.
Indeed, from Euler--Lagrange equations for one independent variable 
($m=1$)
\begin{equation}   \label{euler}
{ \delta L \over \delta q^k } = {\partial L \over \partial q^k  }
- D \left( {\partial L \over \partial \dot{q}^k } \right) = 0,
\qquad
k = 1, ..., n
\end{equation}
we can obtain the canonical Hamiltonian equations (\ref{canonical})
using the Legendre transformation.
It should be noticed that the Legendre transformation is not point.
Hence, there is no conservation of Lie group properties of the
corresponding Euler--Lagrange equations and Hamiltonian equations within
the class of point transformations.

\bigskip

Lie point symmetries in the space $(t,{\bf q},{\bf p})$ are
generated by operators of the form
\begin{equation}  \label{symmetry}
X = \xi ( t, {\bf q}, {\bf p} )  { \partial \over \partial t }
+ \eta ^i  ( t, {\bf q},  {\bf p} )  { \partial \over \partial q ^i }
+  \zeta _i  ( t, {\bf q},  {\bf p})  { \partial \over \partial p _i  }.
\end{equation}
Standard approach to symmetry properties of the Hamiltonian
equations
is to
consider so called {\it Hamiltonian symmetries} \cite{Olver}. In
the case of canonical Hamiltonian equations these are
the evolutionary  ($\xi = 0$) symmetries (\ref{symmetry})
\begin{equation}  \label{evolsymmetry}
X = \eta ^i  ( t, {\bf q},  {\bf p} )  { \partial \over \partial q ^i }
+   \zeta _i  ( t, {\bf q},  {\bf p})  { \partial \over \partial p _i  }
\end{equation}
with
\begin{equation}  \label{integration}
\eta ^i   = {  \partial I  \over \partial p_i  } ,
\qquad
\zeta _i  = -  {  \partial I  \over \partial q^i  },
\qquad
i = 1, ..., n
\end{equation}
for some function $ I (t,{\bf q},{\bf p})$,
namely,  symmetries of the form
\begin{equation}    \label{Hsymmetry}
X_I  =   {  \partial I  \over \partial p_i  }
{ \partial \over \partial q^i  }
-  {  \partial I  \over \partial q^i  }  {
\partial \over \partial p_i  }  .
\end{equation}
These symmetries  are restricted to the phase space
$( {\bf q},{\bf p} )$  and are generated by the function
$I=I(t, {\bf q}, {\bf p})$.
For symmetry (\ref{Hsymmetry}) the independent variable $t$ is
invariant and plays a role of a parameter.

Noether's theorem (Theorem 6.33 in \cite{Olver})
relates Hamiltonian symmetries of the
Hamiltonian equations with their first integrals.
Restricted to the case of the canonical Hamiltonian equations
it can be formulated as follows:

\medskip

\noindent {\bf Proposition.}
{\it An evolutionary vector field $X$ of the form (\ref{evolsymmetry})
generates a Hamiltonian symmetry group of the canonical
Hamiltonian system (\ref{canonical}) if and only if there exists a
first integral  $ I (t,{\bf q},{\bf p})$ so that $ X = X_I $ is
the corresponding Hamiltonian vector field.
Another function $ \tilde{I} (t,{\bf q},{\bf p})$
determines the same Hamiltonian symmetry if and only if $
\tilde{I} = I + F(t) $ for some time-dependent function $ F(t) $.}

\medskip



Indeed, the
invariance of the canonical Hamiltonian equations
(\ref{canonical}) with respect to the symmetry (\ref{Hsymmetry})
leads to
$$
{ \partial ^2  I  \over \partial t \partial p_i }
+ { \partial  H  \over \partial p_j }
{ \partial ^2  I  \over \partial q^j  \partial p_i }
- { \partial  H  \over \partial q^j }
{ \partial ^2  I  \over \partial p_j  \partial p_i }
=  { \partial  I  \over \partial p_j }
{ \partial ^2  H  \over \partial q^j  \partial p_i }
-  { \partial  I  \over \partial q^j }
{ \partial ^2  H  \over \partial p_j  \partial p_i },
\qquad  i = 1, ..., n ,
$$
$$
- { \partial ^2  I  \over \partial t \partial q^i }
- { \partial  H  \over \partial p_j }
{ \partial ^2  I  \over \partial q^j  \partial q^i }
+ { \partial  H  \over \partial q^j }
{ \partial ^2  I  \over \partial p_j  \partial q^i }
= -  { \partial  I  \over \partial p_j }
{ \partial ^2  H  \over \partial q^j  \partial q^i }
+  { \partial  I  \over \partial q^j }
{ \partial ^2  H  \over \partial p_j  \partial q^i },
\qquad  i = 1, ..., n .
$$
These equations can be rewritten as
$$
{ \partial  \over  \partial  p_i }
\left(
{ \partial I  \over \partial t }
+ { \partial  H  \over \partial p_j }
{ \partial   I  \over \partial q^j }
- { \partial  H  \over \partial q^j }
{ \partial  I  \over \partial p_j  }
\right) = 0 ,
\qquad  i = 1, ..., n ,
$$
$$
{ \partial  \over  \partial  q^i }
\left(
{ \partial I  \over \partial t }
+ { \partial  H  \over \partial p_j }
{ \partial   I  \over \partial q^j }
- { \partial  H  \over \partial q^j }
{ \partial  I  \over \partial p_j  }
\right) = 0 ,
\qquad  i = 1, ..., n .
$$
Therefore, we get
$$
{ \partial I  \over \partial t }
+ { \partial  H  \over \partial p_j }
{ \partial   I  \over \partial q^j }
- { \partial  H  \over \partial q^j }
{ \partial  I  \over \partial p_j  }
= f(t)  .
$$
The left hand side stands for the total time derivative of $I$ on
the solutions of the canonical Hamiltonian equations:
$$
 { \partial I  \over \partial t }
+ { \partial  H  \over \partial p_j }
{ \partial   I  \over \partial q^j }
- { \partial  H  \over \partial q^j }
{ \partial  I  \over \partial p_j  }
= \left. D (I) \right|
 _{ \dot{\bf q} = H_{\bf p}, \ \dot{\bf p} = - H_{\bf q} } . 
$$
Thus, we obtain that the Hamiltonian symmetry determines a first
integral of the canonical Hamiltonian equations up to some
time-dependent function, which can be found with the help of these
equations.
This approach has two disadvantages. First, some transformations loose
their geometrical sense if considered in evolutionary form~(\ref{Hsymmetry}).
Second, there is a necessity of integration to find
first integrals with the help of~(\ref{integration}). 
In this approach it is also not clear why
some point symmetries of Hamiltonian equations yield integrals,
while others do not.

In the present paper we will consider symmetries of the general form
(\ref{symmetry}), which are not restricted to the phase space and
can also transform $t$. In contrast to the Hamiltonian symmetries in
the form (\ref{Hsymmetry}) the underlying symmetries have a clear
geometric sense in finite space and do not require integration 
to find first integrals. We will provide a Hamiltonian
version of the Noether's theorem (in the strong formulation)  based
on a newly established Hamiltonian identity, which is an analog of
well-known Noether's identity for Lagrangian approach. The
Hamiltonian identity links directly an invariant Hamiltonian
function with first integrals of the canonical Hamiltonian
equations. This approach provides a simple and clear way to
construct first integrals by means of merely algebraic manipulations
with symmetries of the action functional. The approach will be
illustrated on a number of examples, including equations of the
three-dimensional Kepler motion.

\medskip

The paper is organized as follows: in section~2 we introduce
a definition of an {\it invariant Hamiltonian} and establish the
necessary and sufficient condition for $H$ to be invariant.
 Section~3 contains the main propositions of present
paper: Lemma~1 introduces a new identity, which is used in
Theorem~2 to formulate the necessary and sufficient condition for
existence of first integrals of Hamiltonian equations (Hamiltonian
version of Noether's theorem in the strong formulation). In
section~4 Lemma~2 introduces two more identities, which are used in
Theorem~3 to formulate a sufficient condition for the canonical
Hamiltonian equations to be invariant. Section~\ref{applications}
contains example ODEs  in the form of Euler--Lagrange equations
and Hamiltonian systems. In particular, we consider the
three-dimensional Kepler motion. Final section~\ref{conclusion}
contains concluding remarks.

\section{Invariance of elementary Hamiltonian action }
\label{invariance_section}

As an analog of Lagrangian elementary action \cite{Ibr} we consider
{\it Hamiltonian elementary  action}
\begin{equation}  \label{action}
  p_i  d q^i   -  H dt ,
\end{equation}
which can be invariant or not with respect to a group
generated by an operator of the form (\ref{symmetry}).

\medskip

\noindent {\it Definition.}$\ $
We call a Hamiltonian function invariant with respect to a symmetry
(\ref{symmetry}) if the elementary action (\ref{action}) is an
invariant of the group generated by the operator (\ref{symmetry}).

\medskip

\noindent {\bf Theorem 1.}
A Hamiltonian is invariant with respect to a group with the operator
(\ref{symmetry}) if and only if the following condition holds
\begin{equation}  \label{invar}
    {  \zeta} _i   \dot{q} ^i  +  p _i  D( {  \eta}^i  )
-  X ( H ) - H  D( \xi )  = 0.
\end{equation}

\medskip

\noindent {\it Proof.}$\ $
The invariance condition follows directly from  the action of $X$
prolonged on the differentials  $dt$ and  $d{ q^i }$, $i = 1, ..., n$:
\begin{equation}  \label{prol}
X = \xi ( t, {\bf q}, {\bf p} )  { \partial \over \partial t } +
\eta ^i  ( t, {\bf q},  {\bf p} )  { \partial \over \partial q^i } +
\zeta _i  ( t, {\bf q},  {\bf p})  { \partial \over \partial p_i }+
D(\xi ) dt  { \partial \over \partial (dt) } + D(\eta ^i ) dt { \partial
\over \partial {(dq ^i)} } .
\end{equation}
Application of (\ref{prol}) yields:
$$
X \left(      p_i  d q ^i  - H dt \right) =
\left(   {  \zeta} _i \dot{q} ^i  + p_ i  D( {  \eta}^i  )
-  X ( H ) - H D(\xi ) \right)  dt =  0.
$$
\hfill $\Box$

\medskip

\noindent {\bf Remark 1.}
 From the relation
\begin{equation}  \label{Legen}
 L  ( t, {\bf q} , { \dot{\bf q}} )dt = p_i   d{q} ^i  -
 H ( t, {\bf q} , {\bf p}  ) dt
\end{equation}
it follows  that if a Lagrangian is invariant with respect to a group of
Lie point transformations, then the Hamiltonian
is also invariant with respect to the same group (of point transformations).
The converse statement is false.
For example, symmetries providing components of Runge--Lenz
vector as first integrals of Kepler motion are point symmetries
in Hamiltonian framework (point 5.3).
However, they are generalized symmetries in
Lagrangian framework~\cite{Olver}.

\medskip

\noindent The proof follows from the action of operator
(\ref{symmetry}) on relation (\ref{Legen}).

\medskip

\noindent {\bf Remark 2.} The operator of total differentiation
(\ref{derivative}) applied to Hamiltonian $H$  on the solutions of
Hamiltonian equations (\ref{canonical})  coincides with partial
differentiation with respect to time:
\begin{equation}  \label{diff}
\left. D(H) \right|
 _{ \dot{\bf q} = H_{\bf p}, \ \dot{\bf p} = - H_{\bf q} }
= \left[
      { \partial H \over  \partial t }
+     \dot{q}^i  { \partial H \over \partial q^i  }
+     \dot{p}_ i { \partial H \over \partial p_i  }
\right]
 _{ \dot{\bf q} = H_{\bf p}, \ \dot{\bf p} = - H_{\bf q} }
= { \partial H \over  \partial t }.
\end{equation}

\medskip

\noindent {\bf Remark 3.} The condition (\ref{invar}) means that
$(p_i  d{q}^i   -  H dt)$ is an invariant in space
$({\bf p},{\bf q},d{\bf q},dt)$.  Meanwhile, this condition can be
obtained as an invariance condition of the {\it manifold}
\begin{equation}  \label{h}
h =  p_i  \dot{q}^i  -  H
\end{equation}
under the action of operator (\ref{symmetry}) which is specially
prolonged on the new variable $h$ in the following way:
\begin{equation}  \label{oper}
X = \xi ( t, {\bf q}, {\bf p} )  { \partial \over \partial t }
+  \eta ^i  ( t, {\bf q}, {\bf p})   { \partial \over \partial q^i  }
+  \zeta _i  ( t, {\bf q}, {\bf p} )  { \partial \over \partial p_i  }
- h D(\xi){ \partial \over \partial h }.
\end{equation}
Indeed, application of operator (\ref{oper}) to  (\ref{h}) yields
\begin{equation}
- hD(\xi)=   \zeta _i  \dot{q} ^i
+ p_i  ( D ( \eta ^i )  - \dot{q}_i  D(\xi) ) - X(H) .
\end{equation}
Then,  substitution of $h$ from (\ref{h}) gives the condition
(\ref{invar}).

\section{The Hamiltonian identity and Noether--type theorem}
\label{identity}

Now we can relate conservation properties of the canonical
 Hamiltonian equations to the invariance of the Hamiltonian function.

\medskip

\noindent {\bf Lemma 1.}
The identity
\begin{equation}  \label{ident}
\begin{array}{c}
{\displaystyle
   \zeta _i  \dot{q} ^i  + p _ i   D( \eta  ^i )
-  X ( H ) - H  D( \xi )
=  \xi \left( D(H) - { \partial H \over  \partial t } \right)  } \\
\\
{\displaystyle
-   \eta^i
    \left(  \dot{p}_i  +  { \partial H \over  \partial q  ^i } \right)
+  \zeta _i
    \left(  \dot{q}^i  - { \partial H \over  \partial p _i } \right)
+ D \left[    p _i  \eta ^i  - \xi H  \right]  } \\
\end{array}
\end{equation}
is true for any smooth function $H=H(t,{\bf q},{\bf p} )$.

\medskip

\noindent {\it Proof.}$\ $
The identity can be established by direct calculation.
\hfill $\Box$

\medskip

We call this identity the {\it Hamiltonian identity}. This identity
makes it possible to develop the following
result.

\medskip

\noindent  {\bf Theorem 2.} The canonical Hamiltonian equations
(\ref{canonical}) possess a first integral of the form
\begin{equation}  \label{integral}
I =     p_i  \eta ^i  - \xi H
\end{equation}
if and only if the Hamiltonian function is invariant with
respect to operator (\ref{symmetry}) on the solutions of
canonical equations (\ref{canonical}).

\medskip

\noindent {\it Proof.}$\ $
The result follows from identity (\ref{ident}).
\hfill $\Box$

\medskip

 Theorem 2 corresponds to the strong version of the Noether theorem
 (i.e. necessary and sufficient
 condition) for
invariant Lagrangians and Euler--Lagrange  equations~\cite{Ibr}.

\medskip

\noindent {\bf  Remark.} Theorem~2 can be generalized on
the case of the divergence invariance of the Hamiltonian action
\begin{equation}  \label{divinvaraince2}
    \zeta _i \dot{q} ^i  + p_ i  D( \eta ^i )
-  X ( H ) - H  D ( \xi )  = D( V ),
\end{equation}
where $V  = V ( t,{\bf q} ,{\bf p}) $.
If this condition holds on the solutions
of the canonical Hamiltonian equations (\ref{canonical}), then
there is a first integral
\begin{equation}  \label{integ1}
I =      p_i  \eta ^i  - \xi H - V .
\end{equation}

\section{Invariance of canonical Hamiltonian equations}
\label{invariance}

In the Lagrangian framework, the variational principle
provides us with Euler--Lagrange equations. It is known that the
invariance of the  Euler--Lagrange equations follows from the
invariance of the action integral. The following Lemma~2 and
Theorem 3 establish the sufficient conditions for canonical
Hamiltonian equations to be invariant.

\medskip

\noindent {\bf Lemma 2.} 
The following identities are true for any smooth function
$H=H(t,{\bf q},{\bf p} )$:

$$
{ \delta \over \delta p _j }
\left( {  \zeta} _i   \dot{q} ^i  + p _i  D( {  \eta}^i  )
- X ( H)  - H  D( \xi ) \right)
$$
\begin{equation}  \label{conse1}
\equiv
D ( \eta^j  ) -  \dot{q} ^j  D ( \xi )
- X \left( { \partial H \over \partial p_ j  } \right)
+  { \partial \xi \over \partial p_j }
\left(  D(H) - { \partial H \over  \partial t } \right)
\end{equation}
$$
-    { \partial  \eta ^i  \over \partial p_j }
   \left(  \dot{p}_i  +  { \partial H \over  \partial q ^i  } \right)
+  \left(  { \partial  \zeta _i  \over   \partial  p_j }
    + \delta_{ij} D( \xi ) \right)
   \left( \dot{q}^i  - { \partial H \over  \partial p_i  } \right)  ,
\qquad j = 1, ..., n ,
$$

\medskip

$$
{\delta \over \delta q ^j } \left( {  \zeta} _i   \dot{q} ^i
+  p _i  D( {  \eta}^i  ) - X ( H)  - H  D( \xi ) \right)
$$
\begin{equation}  \label{conse2}
\equiv
 - D ( \zeta _j  ) +  \dot{p} _j  D ( \xi )
 -  X \left( { \partial H \over \partial q_j  } \right)
 +  { \partial \xi  \over \partial q ^j }
   \left( D(H) - { \partial H \over  \partial t }  \right)
\end{equation}
$$
-   \left(  { \partial  \eta ^i \over \partial q ^j }
    +  \delta_{ij} D(\xi) \right)
  \left(  \dot{p}_i  +  { \partial H \over  \partial q ^i  } \right)
+   { \partial  \zeta _i  \over \partial  q ^j }
  \left(  \dot{q} ^i  - { \partial H \over  \partial p_i  }  \right) ,
 \qquad
j = 1, ..., n ,
$$
where the notation $  \delta_{ij}  $ stands for the Kronecker symbol.

\medskip

\noindent {\it Proof.}$\ $
The identities can be easily obtained by direct computation.
\hfill $\Box$

\medskip

\noindent {\bf Theorem 3.} If a Hamiltonian is invariant with
respect to a symmetry (\ref{symmetry}), then Eqs.~(\ref{canonical})
are also invariant.

\medskip

\noindent {\it Proof.}$\ $
   We begin with the invariance of the canonical Hamiltonian
equations (\ref{canonical}). Application of the symmetry operator to
these equations yields
\begin{equation}  \label{cons1}
D ( \eta^j  ) -  \dot{q} ^j  D ( \xi )
- X \left( { \partial H \over \partial p_ j  } \right) = 0 ,
\qquad
j = 1, ..., n ;
\end{equation}
\begin{equation}  \label{cons2}
D ( \zeta _j  ) -  \dot{p} _j  D ( \xi )
+ X \left( { \partial H \over \partial q ^j  } \right) = 0,
\qquad
j = 1, ..., n .
\end{equation}
Both conditions  obtained should be true on the solutions of
Eqs.~(\ref{canonical}).

Let  Hamiltonian be invariant, then  all left hand sides of
identities (\ref{conse1}) and  (\ref{conse2}) are equal to zero on
the solutions of Eqs.~(\ref{canonical}).  All right hand sides of
(\ref{conse1}) and  (\ref{conse2}) are also equal zero. Substituting
equations (\ref{canonical}) into the right hand sides of
(\ref{conse1}) and  (\ref{conse2}), we obtain the invariance
conditions (\ref{cons1}),(\ref{cons2}). \hfill $\Box$

\medskip

\noindent {\bf Remark.} The statement of Theorem~3
remains valid if we consider divergence symmetries of
the Hamiltonian, i.e. condition (\ref{divinvaraince2}),
because the right hand side term $D(V)$ belongs to the kernel of the
variational operators (\ref{varoperator1}),(\ref{varoperator2}).

\medskip

The invariance of the Hamiltonian on equations (\ref{canonical})
is a {\it sufficient condition} for the canonical Hamiltonian
equations to be invariant. 
The following Theorem 4 establishes the {\it necessary and
sufficient} condition  for  canonical Hamiltonian equations
to be invariant.

\medskip

\noindent {\bf Theorem 4.} Canonical Hamiltonian equations
(\ref{canonical}) are  invariant with respect to a symmetry
(\ref{symmetry}) {if and only if} the following conditions are
true (on the solutions of  the canonical Hamiltonian equations):
\begin{equation}  \label{new1}
\begin{array}{l}
{\displaystyle
\left. { \delta \over \delta p _j } \left( {  \zeta} _i   \dot{q} ^i
+ p _i  D( {  \eta}^i  ) - X ( H)  - H  D( \xi )
\right) \right|
 _{ \dot{\bf q} = H_{\bf p}, \ \dot{\bf p} = - H_{\bf q} }
=0 ,
\qquad j = 1, ..., n ;  } \\
\\
{\displaystyle
\left. {\delta \over \delta q ^j } \left( {  \zeta} _i   \dot{q} ^i
+  p _i  D( {  \eta}^i  ) - X ( H)  - H  D( \xi )
\right)  \right|
 _{ \dot{\bf q} = H_{\bf p}, \ \dot{\bf p} = - H_{\bf q} }
=0  ,
\qquad j = 1, ..., n . } \\
\end{array}
\end{equation}

\medskip

\noindent {\it Proof.}$\ $  The proof follows from identities
(\ref{conse1}) and (\ref{conse2}).
\hfill $\Box$

\medskip

It should be noted that conditions (\ref{new1}) are true for all
symmetries of canonical Hamiltonian equations. But not all of those
symmetries yield the "variational integral" of these conditions, i.e.
$$
 \left( {  \zeta} _ i   \dot{q} ^i  + p
_i  D( {  \eta}^i  ) - X ( H)  - H  D( \xi )
\right)|
 _{ \dot{\bf q} = H_{\bf p}, \ \dot{\bf p} = - H_{\bf q} } 
=0,
$$
which gives first integrals  due to Theorem 2. {That is why not all
symmetries of the canonical Hamiltonian equations provide first
integrals}. We will illustrate all theorems, given above, on the
following examples.

\section{Applications}
\label{applications}

In this section we provide examples how to find first integrals
using symmetries.

\subsection{Example 1. A scalar ODE. }

 As the first example we consider the following second-order ODE
\begin{equation} \label{equ1}
\ddot{u} = \frac{1}{u^{3}},
\end{equation}
which admits Lie algebra $L_3$ with basis operators
\begin{equation} \label{ope3}
 X_1 = \dx t , \quad X_2 = 2t \dx t + u\dx u ,  \quad
X_3 =t^2 \dx t + tu \dx u.
\end{equation}

\subsubsection{Lagrangian approach}

The Lagrangian function
\begin{equation} \label{L1}
L ( t, u, \dot{u})  =  {1 \over 2} \left(  \dot{u}^2 - \frac{1}{u^2} \right)
\end{equation}
is invariant with respect to  $X_1, X_2$. Therefore, by means of
 Noether's theorem there exist the following first integrals
\begin{equation} \label{int2}
J_{1} =   - {1 \over 2} \left(  \dot{u}^2 + \frac{1}{u^2} \right) ,
\qquad
J_{2} =  u  \dot{u} -  t \left(  \dot{u}^2 + \frac{1}{u^2} \right) .
\end{equation}

The action of the third operator $X_3$ yields a divergence
invariance condition
\begin{equation} \label{div2}
X {L} +  {L} D (\xi)  =  u  \dot{u} = D \left( { u^2 \over 2}  \right).
\end{equation}
Due to the divergence invariance of the Lagrangian
we can find the following first integral
\begin{equation} \label{int3}
J_{3} =  - { 1 \over 2}
\left( {t^2 \over u^2} + ( u - t \dot{u} ) ^2 \right) .
\end{equation}
Alternatively, one can find the last integral from another
Lagrangian function
\begin{equation} \label{L2}
\tilde{L} ( t, u, \dot{u} )  = \left( \frac{u}{t}-  \dot{u}  \right) ^2 - \frac{1}{u^2},
\end{equation}
which is exactly invariant with respect to  $X_3$.

It should be mentioned that independence of first integrals
obtained with the help of the Noether theorem
is guarantied only in the case when there is
one Lagrangian which is invariant with respect to all symmetries. This
condition is broken in the considered example.
Therefore, the integrals obtained are not independent.
Integrals (\ref{int2}),(\ref{int3}) are connected
by the following relations
\begin{equation} \label{connect}
4 J_{1} J_{3} - J_{2} ^2 = 1.
\end{equation}
Thus, any two integrals among (\ref{int2}),(\ref{int3}) are
independent. Putting  $J_1 = A/2 $, $J_2 = B $, and excluding $\dot{u}$ we
find the general solution of equation (\ref{equ1}) as
\begin{equation} \label{sol3}
A u^{2} +  ( A t + B )^{2} + 1 = 0 .
\end{equation}

\subsubsection{Hamiltonian framework}

Let us transfer the preceding example into the Hamiltonian framework.
We change variables
$$
q = u,
\qquad
p = { \partial {L} \over  \partial  \dot{u}  }=  \dot{u}  .
$$
The corresponding Hamiltonian is
\begin{equation} \label{fun}
H (t,q,p) =  \dot{u} { \partial {L} \over  \partial \dot{u} } - {L}
=  { 1 \over 2} \left(  p^2  + { 1 \over  q^2 } \right)  .
\end{equation}

The Hamiltonian equations
\begin{equation}  \label{ca}
\dot{q}  =  p  ,
\qquad
\dot{p} = \frac{1}{q^3}
\end{equation}
admit symmetries
\begin{equation} \label{opera}
 X_1 = \dx t ,
 \qquad
 X_2 = 2t \dx t + q \dx q  -p \dx p ,
 \qquad
 X_3 =t^2 \dx t + tq \dx q  + ( q - tp) \dx p .
\end{equation}

We check invariance of $H$ in accordance with Theorem 1 and find
that condition (\ref{invar}) is satisfied  for the operators $X_1$ and $X_2$.
Then, using Theorem~2,  we calculate the corresponding
first integrals
\begin{equation} \label{integ2}
I_{1} = - H  =  -  { 1 \over 2} \left(  p^2  + { 1 \over  q^2 } \right) ,
\qquad
I_{2} = pq -  t \left(  {p^2}   + { 1 \over  q^2 } \right).
\end{equation}
For the third symmetry operator the Hamiltonian is divergence
invariant with $V_3 = q^2 / 2 $. In accordance with Remark for 
Theorem~2 it yields the following conserved quantity
\begin{equation} \label{integral3}
I_{3} =  - { 1 \over 2}
\left( {t^2 \over q^2} + ( q - t p  ) ^2 \right) .
\end{equation}
Note that no integration is needed.
As we noted before only two first integrals of a second order ODE
can be functionally independent.
Putting  $I_1 = A/2$ and $I_2 = B $,
we  find the solution of equations (\ref{ca}) as
\begin{equation} \label{sol3h}
A q^{2}  +  ( A t -  B  )^{2} + 1  = 0 ,
\qquad
p = { B -  At \over q}.
\end{equation}

\subsubsection{Evolutionary vector field approach}

Let us consider the same example for evolutionary vector fields in the
Hamiltonian form (\ref{Hsymmetry}).
We rewrite operators (\ref{opera})
in the evolutionary form
\begin{equation} \label{operat}
\begin{array}{c}
{\displaystyle
 \bar{X}_1 = - \dot{q}\dx q  -  \dot{p}\dx p ,
\quad
 \bar{X}_2 =  (q-2t\dot{q})\dx q  -(p+2t\dot{p})\dx p ,} \\
\\
{\displaystyle
 \bar{X}_3 = (tq-t^2\dot{q}) \dx q  + ( q - tp-t^2\dot{p})\dx p  . } \\
\end{array}
\end{equation}
Transformations which correspond to the symmetries (\ref{operat})
are not point. Therefore,   Hamiltonian equations (\ref{ca}) are
invariant with respect to (\ref{operat}) if being considered
together with their differential  consequences.
On the solutions of the canonical equations (\ref{ca})
these operators are equivalent to the set
\begin{equation} \label{operat2}
\begin{array}{c}
{\displaystyle
 \tilde{X}_1 =    -  p  \dx q  -  \frac{1}{q^3}  \dx p ,
\quad
 \tilde{X}_2 =  (q -  2 t p  )\dx q
-  \left( p + \frac{2t}{q^3}  \right) \dx p ,} \\
\\
{\displaystyle
\tilde{X}_3 = \left( tq - { t^2 } p \right) \dx q
+ \left(  q - tp - \frac{t^2}{q^3} \right) \dx p  . } \\
\end{array}
\end{equation}
One should
integrate the equations (\ref{integration}) for each operator
to find first integrals.  Integration
provides us with three first integrals given in
(\ref{integ2}),(\ref{integral3}).

\subsection{Example 2. Repulsive one-dimensional motion.}

As another example of an ODE we consider one-dimensional motion in
the Coulomb field (the case of a repulsive force):
\begin{equation} \label{equ2}
 \ddot{u}  = \frac{1}{u^{2}},
\end{equation}
which admits Lie algebra $L_2$ with basis operators
\begin{equation} \label{opera3}
 X_1 = \dx t ,
\quad
X_2 = 3t \dx t + 2u\dx u .
\end{equation}

\subsubsection{Lagrangian approach}

The Lagrangian function
\begin{equation} \label{Lag}
L ( t, u, \dot{u} )  = { \dot{u} ^2 \over 2 }  - \frac{1}{u}
\end{equation}
is invariant only with respect to  $X_1$.
Therefore,  Noether's theorem yields the only first integral
\begin{equation} \label{inte}
J_{1} =  { \dot{u} ^2 \over 2 }  + { 1 \over  u }.
\end{equation}
In this case the Euler--Lagrange equation admits two symmetries
while the Lagrangian is invariant  with respect to one symmetry
operator only.

\subsubsection{Hamiltonian framework}

We change variables
$$
q=u,
\qquad
p={ \partial {L} \over  \partial \dot{u} }= \dot{u}
$$
and find the Hamiltonian function
\begin{equation} \label{hamilt}
H (t, q, p) =  \dot{u} { \partial {L} \over  \partial \dot{u} } - {L}
=  \frac{p^2}{2} + { 1 \over  q } .
\end{equation}

The Hamiltonian equations have the form
\begin{equation}  \label{candy}
\dot{q} =  p ,
\qquad
\dot{p} = \frac{1}{q^2} .
\end{equation}
We rewrite symmetries in the canonical variables as the following
algebra $L_2$
\begin{equation} \label{operato}
 X_1 = \dx t ,
\quad
X_2 = 3t \dx t + 2q \dx q  - p  \dx p .
\end{equation}
The invariance of Hamiltonian condition (\ref{invar}) is satisfied
for the operator $X_1$ only. Applying Theorem 2, we calculate the
corresponding first integral
\begin{equation} \label{integ3}
I_{1} =   - \left(  \frac{p^2}{2} + { 1 \over  q } \right) .
\end{equation}
Application of operator $X_2$ to the Hamiltonian action gives
\begin{equation} \label{qu}
 {  \zeta}   \dot{q}   + p  D( {  \eta} ) - X  ( H) -
H  D( \xi ) =p \dot{q} -\left( \frac{p^2}{2} + { 1 \over  q
}\right)\neq 0.
\end{equation}
Meanwhile,  in accordance with Theorem 4 we have
$$
\left.
{ \delta \over \delta p } \left( {  \zeta}   \dot{q}  + p
D( {  \eta}  ) - X ( H)  - H  D( \xi )
\right)
\right|
_{ \dot{q} =  p , \ \dot{p} = \frac{1}{q^2}  }
= ( \dot{q} - p ) |
_{ \dot{q} =  p , \ \dot{p} = \frac{1}{q^2}  }
= 0 ,
$$
$$
\left.
{\delta \over \delta q } \left( {  \zeta}  \dot{q}  +  p
  D( {  \eta}  ) - X ( H)  - H  D( \xi )
\right)
\right|
_{ \dot{q} =  p , \ \dot{p} = \frac{1}{q^2}  }
= \left. \left(  - \dot{p} + { 1 \over q^2 } \right)
\right|
_{ \dot{q} =  p , \ \dot{p} = \frac{1}{q^2}  }
=0.
$$
We will show below that there exists a second integral of non-local
character.

\bigskip

It was shown in \cite{Dor} that Eq.~(\ref{equ2}) can be linearized by
a contact transformation. For  equations (\ref{candy}) this
transformation is the following
\begin{equation} \label{change}
p(t) = P(s) ,  \quad Q^2(s)= \frac{2}{q(t)} ,
 \quad  dt = -
\frac{4}{Q^3} ds.
\end{equation}
The new Hamiltonian
\begin{equation} \label{hamilton}
H (s, Q, P)  =  { 1 \over 2 } ( {P^2} + { Q^2} )
\end{equation}
corresponds to the linear equations
\begin{equation}  \label{canHam}
{ dQ \over ds } =  P   , \qquad { dP \over ds } = - Q,
\end{equation}
which describe the one-dimensional harmonic oscillator.
These  equations have  two first  integrals
\begin{equation} \label{integrals}
\tilde{I}_{1} =  {1 \over 2 } ( {P^2} + {Q^2 } )  ,
\qquad
\tilde{I}_{2} = \arctan \left( \frac{P}{Q}\right) + s ,
\end{equation}
which let us write down the general solution of equations (\ref{canHam}) as 
\begin{equation} \label{sine}
Q =  A \sin s + B \cos s, \qquad P= A \cos s - B \sin s,
\end{equation}
where $A$ and $B$ are arbitrary constants. 
Applying the transformation (\ref{change}) to integral $I_{2}$ one
can develop the  non-local integral for Eqs.~(\ref{candy})
\begin{equation} \label{integra}
I_2 ^* = \arctan \left( \frac{p\sqrt q}{\sqrt 2}\right) -
   \frac{1}{\sqrt 2}  \int_{t_0}^t \frac{dt}{ q^{3/2} }.
\end{equation}

\subsection{Example 3. Three-dimensional Kepler motion}

The three-dimensional Kepler motion of a body in Newton's
gravitational field is given by the equations
\begin{equation}  \label{Kepler}
 { d {\bf q} \over dt}  =  {\bf p}  , \qquad
 { d {\bf p} \over dt } =
- { K ^2 \over r  ^3  }  {\bf q} ,
\qquad r = | {\bf q} | ,
\end{equation}
where $ {\bf q} = ( q_1, q_2, q_3) $, $ {\bf p} = ( p_1, p_2, p_3) $
and $K$ is a constant,  with the initial data
$$
{\bf q} ( 0 ) = {\bf q} _0 , \qquad
{\bf p} ( 0 ) = {\bf p} _0  .
$$
These equations are Hamiltonian. They are defined by the following
Hamiltonian function
\begin{equation}  \label{hamil}
H ( { \bf  q }, {\bf p } )
= { 1  \over 2 }  | { \bf  p } | ^2  -  { K ^2 \over r   }  .
\end{equation}

Among symmetries admitted by the equations (\ref{Kepler})
there are
$$
\begin{array}{l}
{\displaystyle
X_0 = { \partial \over \partial t} ,
\qquad
X_1 = 3t  { \partial \over \partial t}
+ 2 q_i { \partial \over \partial q_i}
- p_i { \partial \over \partial p_i}  ,} \\
\\
{\displaystyle
X_{ij} = - q_j { \partial \over \partial q_i}
+  q_i { \partial \over \partial q_j}
- p_j { \partial \over \partial p_i}
+  p_i { \partial \over \partial p_j} ,
\qquad   i \neq j    ,} \\
\\
{\displaystyle
Y_l =  ( 2 q_l p_k - q_k p_l - ( {\bf q},  {\bf p} ) \delta _{lk} )
 { \partial \over \partial q_k}  } \\
\\
{\displaystyle
+ \left(
   p_l p_k -  ( {\bf p},  {\bf p} ) \delta _{lk}
- { K^2 \over r^3 }  (  q_l q_k -  ( {\bf q},  {\bf q} ) \delta _{lk}  )
  \right)
{ \partial \over \partial p_k} ,
\qquad
 l = 1,2,3 , } \\
\end{array}
$$
where $( {\bf f},  {\bf g} ) = {\bf f}^T {\bf g} $
is scalar product of vectors.

The Hamiltonian function (\ref{hamil}) is invariant for symmetries
$X_0$ and $X_{ij} $. Theorem~2 makes it possible
 we find the first integral for symmetry $X_0$
$$
I_1 = - H ,
$$
which represents the conservation of energy in Kepler motion.
For symmetries  $X_{ij}$ we obtain the first integrals
$$
I_{ij} = q_i p_j  -  q_j p_i , \qquad i \neq j ,
$$
which are components of the angular momentum
\begin{equation}  \label{moment}
{\bf L} ( { \bf  q }, {\bf p } )  = {\bf q} \times {\bf p}  .
\end{equation}
Conservation of the angular momentum shows that
the orbit of motion of a body lies in a fixed plane
perpendicular to the constant vector $ {\bf L} $.
It also follows that in this plane the position
vector $  {\bf q} $ sweeps out equal areas in equal time intervals,
so that the sectorial velocity is constant  \cite{Arnold}.
Therefore, Kepler's second law can be considered as a geometric
reformulation of the conservation of angular momentum.

The scaling symmetry $ X_1$ is not a Noether symmetry
(neither variational, nor divergence symmetry)
and does not lead to a conserved quantity.

For each of symmetries    $Y_{l}$ the Hamiltonian is
divergence invariant with functions
$$
V_l = q_l  \left(  ( {\bf p},  {\bf p} ) + { K^2 \over r } \right)
- p_l  ( {\bf q},  {\bf p} )  , \qquad  l  = 1, 2, 3.
$$
Hence, the operators   $Y_{l}$  yield the first integrals
$$
I_{l} = q_l  \left(  ( {\bf p},  {\bf p} ) - { K^2 \over r } \right)
- p_l  ( {\bf q},  {\bf p} ) ,
\qquad  l = 1, 2, 3,
$$
which are components of the Runge--Lenz vector
\begin{equation}
{ \bf A}  ( { \bf  q }, {\bf p } ) =
{\bf p} \times {\bf L} - { K^2   \over r }  {\bf q} =
{ \bf q }  \left(  H ( { \bf  q }, {\bf p } )
+  { 1 \over 2 }  | {\bf p } | ^2   \right)
- { \bf p } ( {\bf q } ,  {\bf p } )  .
\end{equation}
Physically, vector ${ \bf A}$ points along the major axis of
the conic section determined by the orbit of the body. Its
magnitude determines the eccentricity \cite{Thirring}.

Let us note that not all first integrals are independent. There
are two relations between them given by the equations
$$
 { \bf A} ^2 - 2 H  {\bf L}^2  =  K ^4  \qquad
\mbox{and}
\qquad
(  { \bf A}  ,  { \bf L } )  = 0 .
$$

\bigskip

The two-dimensional Kepler motion can be considered in a similar way.
Let us remark that symmetries and first integrals
of the two-dimensional Kepler
motion can be obtained by restricting the symmetries and first integrals of the
three-dimensional Kepler motion to the space $(t, q_1, q_2, p_1, p_2)$.
As the conserved quantities of the two-dimensional Kepler
motion one obtains the energy
$$
H ( { \bf  q }, {\bf p } )
= { 1  \over 2 }  | { \bf  p } | ^2  -  { K ^2 \over r   },
\qquad r = | { \bf  q } | ,
\qquad { \bf  q } = ( q_1, q_2 ) ,
\quad {\bf p } = ( p_1, p_2 ) ,
$$
one component of the angular momentum
$$
L_3  = { q_1 } { p_2 }  -  { q_2 } { p_1 }
$$
and
two components of the  Runge--Lenz vector
$$
A_1 = q_1  \left(  H ( { \bf  q }, {\bf p } )
+  { 1 \over 2 }  | {\bf p } | ^2   \right)
-  p_1  ( {\bf q } ,  {\bf p } )  .
$$
$$
A_2 = q_2  \left(  H ( { \bf  q }, {\bf p } )
+  { 1 \over 2 }  | {\bf p } | ^2   \right)
-  p _2  ( {\bf q } ,  {\bf p } )  .
$$
There is one relation between these conserved quantities, namely
$$
A_1 ^2 +  A_2  ^2 - 2 H  L_3 ^2  =  K ^4   .
$$

\bigskip

Further restriction to the one-dimensional Kepler motion leaves only one
first integral, which is the Hamiltonian function.

\section{Conclusion}
\label{conclusion}

In the present paper we considered how invariance of the action
functional can be related to first integrals of  canonical
Hamiltonian equations. The conservation properties of the canonical
Hamiltonian equations were based on the newly  established identity,
which was called the Hamiltonian identity, and which can be viewed
as a "translation" of the well-known Noether identity into
Hamiltonian approach. This identity makes it possible to establish
one-to-one correspondence between invariance of the Hamiltonian and
first integrals of the canonical Hamiltonian equations (the strong
version of Noether's theorem).

The variational consequences of the Hamiltonian identity make it
possible to establish  necessary and sufficient conditions for the
canonical Hamiltonian equations to be invariant. This makes clear
why not each symmetry of the Hamiltonian equations provides a first
integral.

The Hamiltonian version of Noether's theorem,
formulated in the paper, gives  a
constructive way to find first integrals of the canonical Hamiltonian
equations once their symmetries are known. This simple method does
not require integration as it was illustrated by a number of examples. In
particular, we considered equations of Kepler motion in various
dimensions. The presented  approach gives a possibility to consider
canonical Hamiltonian equations and find their first integrals
without exploiting the relationship to the Lagrangian formulation
(see, for example, \cite{Str}).

The discrete and semi-discrete versions of presented constructions
will be considered elsewhere.

\bigskip
\bigskip

\noindent{\bf \large Acknowledgments}

\medskip

The V.D.'s research was sponsored in part by the Russian Fund for
Basic Research under the research project no.~06-01-00707.
The research of R.K. was partly supported by  the Norwegian Research Council
under contract no.~176891/V30.


\end{document}